\documentclass[11pt]{article}
 \usepackage{amsfonts}
 \usepackage{amssymb}
 \def\bsh{\backslash}

 \newfont{\bbbold}{msbm10 scaled \magstep1}

 \def\bbC{\mbox{\bbbold C}}

 \def\bbF{\mbox{\bbbold F}}

 \def\bbR{\mbox{\bbbold R}}

 \def\cL{{\cal L}}
 
 \def\cN{{\cal N}}

 \def\cV{{\cal V}}

 \newfont{\goth}{eufm10 scaled \magstep1}

 \def\gs{\mbox{\goth s}}
 
 \def\gu{\mbox{\goth u}}

 \def\a{\alpha}
 \def\b{\beta}
 \def\c{\gamma}
 \def\d{\delta}\def\D{\Delta}
 \def\e{\epsilon}
 \def\vf{\varphi}
 \def\h{\eta}

 \def\l{\lambda}

 \def\th{\theta}
 
 \def\be{\begin{equation}}\def\ee{\end{equation}}
 \def\bea{\begin{eqnarray}}\def\eea{\end{eqnarray}}
 \def\ba{\begin{array}}\def\ea{\end{array}}

 \def\del{\partial}

 \def\xz{\times}

 \def\del{\partial}

 \let\la=\label

 {}

 \def\nn{\nonumber}
 \def\bd{\begin{document}}
 \def\ed{\end{document}}
 \def\bea{\begin{eqnarray}}\def\barr{\begin{array}}\def\earr{\end{array}}
 \def\eea{\end{eqnarray}}
 \def\ft#1#2{{\textstyle{{\scriptstyle #1}\over {\scriptstyle #2}}}}
 \def\fft#1#2{{#1 \over #2}}
 \newcommand{\eq}[1]{(\ref{#1})}
 \def\eqs#1#2{(\ref{#1}-\ref{#2})}
 \def\det{{\rm det\,}}
 \def\tr{{\rm tr}}\def\Tr{{\rm Tr}}

\begin{document}

 \thispagestyle{empty}

 \hfill{KCL-TH-00-49}

 \hfill{hep-th/0008048}

 \hfill{\today}

 \vspace{20pt}

 \begin{center}
 {\Large{\bf Aspects of the $D=6$, $(2,0)$ Tensor Multiplet}}
 \vspace{30pt}

 {P.S. Howe} \vskip 1cm {Department of Mathematics}
 \vskip 1cm {King's College, London} \vspace{15pt}

 \end{center}

 \vspace{60pt}

 {\bf Abstract}

Some aspects of the $D=6, (2,0)$ tensor multiplet are discussed.
Its formulation as an analytic superfield on a suitably defined
superspace and its superconformal properties are reviewed. Powers
of the field strength superfield define a series of superconformal
fields which correspond to the KK multiplets of $D=11$ supergravity
on an $AdS_7\xz S^4$ background. Correlation functions of these
operators are briefly discussed.

 {\vfill\leftline{}\vfill \vskip  10pt

 \baselineskip=15pt \pagebreak \setcounter{page}{1}

 \section{Introduction}

 The $D=6, (2,0)$ tensor multiplet plays an important r\^{o}le in
 modern string theory through the fact that it is the world-volume
 multiplet of the M-theory 5-brane. In the context of the Maldacena
 conjecture \cite{mal} the near-horizon geometry of a large number of such
 branes is $AdS_7\xz S^4$, and $D=11$ supergravity on such a background is
 supposed to be related to a conformal field theory of the boundary
 of $AdS_7$ \cite{d6}. In the $AdS_5\xz S^5$ case for IIB supergravity, the
 boundary conformal field theory is $N=4$ super Yang-Mills theory,
 in the large $N_c$ limit (where the gauge group is $SU(N_c)$), but
 in the $D=6$ case the interacting conformal field theory is not
 known. Nevertheless some progress can be made assuming that there
 is such a superconformal field theory.

 The $D=6, (2,0)$ tensor multiplet was first written down in
 \cite{hsit} and reformulated in an appropriate harmonic superspace 
 in  \cite{h}
 where it was also shown how the superconformal fields which
 correspond to the KK multiplets of $D=11$ supergravity can be
 easily constructed as powers of the underlying field strength
 superfield. This construction is very similar to the
 construction of the KK multiplets
 which was used for $N=4$ Yang-Mills theory in \cite{hw, af}
 following the realisation that the field strength tensor and the
 supercurrent can be written in a simple way in superspace \cite{hh2}.

 In this paper we give more details of the results outlined in \cite{h};
 after discussing the multiplet in ordinary superspace\footnote{More recent discussions of the multiplet in ordinary superspace can be found in \cite{kvp,gm}.}, we
 reformulate it in harmonic superspace. We then go on to discuss
 its conformal properties and finally we make some remarks on the
 properties of correlation functions that can be deduced from
 symmetry properties alone.


 \section{The tensor multiplet}

The (2,0) tensor multiplet consists of five scalars, a set of four
chiral fermions transforming under the four-dimensional
representation of the internal symmetry group $USp(4)$ (symplectic
Majorana-Weyl fermions) and a two-form gauge potential $A$ whose
three-from field strength is self-dual. The coordinates of $D=6,
(n,0)$ Minkowski superspace are $(x^a,\th^{\a i})$ where $a=0,\ldots
5$ is a Lorentz vector index, $\a=1,\ldots 4$ is a chiral spinor
index and $i=1,\ldots 2n$ is an internal symmetry index for the
group $USp(2n)$. The spinors satisfy a reality condition of the form

\be
\psi_{\a i}\mapsto \bar\psi_{\a}^i=\h^{ij}\psi_{\a j} \ee

where the conjugation denoted by the bar includes multiplication by
a matrix $B$ which converts dotted to undotted spinor indices and
satisfies $\bar B B=-1$, so that they have altogether $4n$ real
components. The antisymmetric matrix $\h^{ij}$ is the symplectic
invariant and we use the convention that $\h_{ij}$ has the same
components as $\h^{ij}$. The superspace covariant derivatives are
$\del_a$ and $D_{\a i}$ where

\be
[D_{\a i}, D_{\b j}]=i\h_{ij}(\c^a)_{\a\b}\del_a \ee

The (2,0) tensor multiplet is described by a superfield $W_{ij}$
which is real, antisymmetric and traceless with respect to
$\h^{ij}$. It obeys the constraint

\be
D_{\a i}W_{jk}=2\h_{ij}\l_{\a k}-2\h_{ik}\l_{\a j} + \h_{jk} \l_{\a
i} 
\la{3}
\ee

from which it follows that

\be
D_{\a i}\l_{\b j}=-{i\over2}(\c^a)_{\a\b}\del_a W_{ij} +
{i\over24}(\c^{abc})_{\a\b}F_{abc} \ee

where $F_{abc}$ is self-dual. By applying further $D$'s to these
relations one finds that the leading components of $W_{ij}$,
$\l_{\a i}$ and $F_{abc}$ are the only independent spacetime fields
and that they obey the equations of motion

\bea 
\del^a\del_a W_{ij}&=& (\c^a)^{\a\b}\del_a\l_{\b i}=0 \nn \\
\del^aF_{abc}&=&\del_{[a}F_{bcd]}=0 \eea

The multiplet can also be described by a superspace three-form
$F=dA$ where $A$ is a potential two-form. The non-vanishing
components of $F$ are

\bea F_{\a i\b j c}&=& -i(\c_c)_{\a\b} W_{ij} \nn \\ F_{ab \c k}&=&
(\c_{ab}\l)_{\c k} \la{3f}
 \eea

as well as $F_{abc}$.

We note that there is also a $(1,0)$ tensor multiplet described by
a real scalar superfield $W$ which is subject to a second-order
constraint

\be
D_{[\a}^{(i} D_{\b]}^{j)}W=0 \ee

Its components are a scalar, a spinor $\l_{\a i}$ and a self-dual
field strength $F_{abc}$. It can also be described by a $(1,0)$
superspace three-form $F$ whose components are the same as in
\eq{3f} with $W_{ij}$ replaced by $\h_{ij}W$. (In $(1,0)$
$\h_{ij}=\e_{ij}$.)


\section{Harmonic superspace description}

Harmonic superspaces are extensions of Minkowski superspace $M$ of
the form $\hat M:=M\xz \bbF$ where $\bbF$ is a coset space of the
internal symmetry group which is usually chosen to be a compact
complex manifold \cite{gikos1}. Such superspaces can be thought of as
applications of twistor geometrical techniques in a supersymmetric
setting. The original idea of Penrose's twistor theory was to
encode information about spacetime in holomorphic data on
projective twistor space \cite{penrose}. However, in the case of Euclidean
``spacetime'' $\bbR^4$, twistor space can be thought of as an
extension of spacetime by a two-sphere $S^2\sim \bbC P^1$ \cite{ward}. The
significance of this sphere is that it parametrises the self-dual
complex structures on $\bbR^4$ (there is also an anti-self-dual
set). One can then construct a natural complex structure on the
total space by combining the complex structure on $\bbR^4$
parametrised by the given point of the sphere with the complex
structure on the sphere itself. This idea can be generalised to
superspace, but with complex structures replaced by (odd) CR structures
which can be thought of as  partial complex structures \cite{rs}. More
formally, a CR structure is a subbundle of the complexified tangent
bundle which has local bases of complex vector fields whose Lie brackets are fields of the same type. The prototype of
an odd CR structure is provided by chirality in $D=4, N=1$
superspace and generalised chirality is referred to as Grassmann
analyticity (G-analyticity). In the harmonic superspace context the space
$\bbF$ parametrises the odd, Lorentz covariant CR structures on
Minkowski superspace, and on the total space $\hat M$ there is a
natural CR structure obtained by combining the odd CR structure specified by
a given point of $\bbF$ with the complex structure of $\bbF$ itself.

In $D=6, (n,0)$ superspace the problem is therefore to find subsets
of the $D$'s which anticommute amongst themselves. If we demand
that such subsets be Lorentz covariant then we are looking for
derivatives $D_{\a r}=u_r{}^i D_{\a i}$ such that

\be
[D_{\a r}, D_{\b s}]=0 \la{cr} \ee

since in such a case it is consistent to look for superfields which
are annihilated by $D_{\a r}$. Clearly there will be a
complementary set $D_{\a r'}=u_{r'}{}^i D_{\a i}$, and we should
require that $u_I{}^i=(u_r{}^i,u_{r'}{}^i)$ be an element of
$USp(2n)$. In order to have such odd CR structures we must have

\be
u_r{}^i u_s{}^j \h_{ij}=0 \ee

For $(1,0)$ the only possibility is to choose $r$ to have only one
value, $1$ say, so that the space of odd CR structures is again the
two-sphere. In the $(2,0)$ case a natural choice is to allow $r$ to
take two values, $1,2$, say, with $r'=3,4$. Then we can satisfy
\eq{cr} above in this way if we choose the basis in which

\bea \h_{rs}&=&\h_{r's'}=0 \nn \\ \h_{rs'}&=-&\h_{s'r}=\d_{rs'}
\eea

The space of such odd CR structures is $U(2)\bsh USp(4)$. Clearly
we can generalise this particular type of structure to arbitrary
$n$ with associated coset space $U(n)\bsh USp(2n)$.

There are other possibilities which do not halve the number of odd coordinates; for example, in $(2,0)$ superspace we can consider the CR-structure specified by $D_{\a 1}$ which again has isotropy group $U(1)\xz SU(2)$ (this is not the same group as that for $D_{\a 1}, D_{\a 2}$). Moreover, one can consider Lorentz covariant odd CR structures spanned by $p$ derivatives compatible with odd CR structures spanned by $q$ derivatives, $p<q$, and so on. The spaces associated with such sets of CR structures are clearly the spaces of nested subspaces of $\bbC^{2n}$ which are isotropic, i.e. for any two vectors $u,v$ in a given subspace one has $\h_{ij} u^i v^j=0$. These spaces are called isotropic flag manifolds and have been employed in the recent work of Ferrara and Sokatchev on conformal fields in six dimensions \cite{fs}. In the following we shall only use spaces of the type $U(n)\bsh USp(2n)$.

The required harmonic superspaces are formed by adjoining these
parameter spaces to Minkowski superspace, the idea being that
constraints on Minkowski space superfields can be expressed more
simply in terms of fields defined on the larger spaces. We
therefore form the spaces $\hat M:=M\xz \bbF$ where $\bbF=U(n)\bsh
USp(2n)$. Since $F$ is a homogeneous space it is often convenient
to use the well-known group-theoretical technique of writing tensor
fields on $\bbF$ as equivariant maps from $USp(2n)$ to $V$ where $V$
is a representation space for the isotropy subgroup $U(n)$. Such a
map $A(u)$ has the property that $A(hu)=R(h)A(u)$ where $h\in U(n)$
and $R(h)$ is the representation of $U(n)$ acting on $V$.

An element $u$ of $USp(2n)$ is both unitary and symplectic, so that

\be
u_I{}^i u_J{}^j\h_{ij}=\h_{IJ} \ee

and

\be
\bar u^I{}_i=(u^{-1})_i{}^I:=u_i{}^I \ee

In our conventions the group $USp(2n)$ acts naturally to the right
on $\bbF$, and this right action is generated by the left-invariant
vector fields. On the other hand, the left action of the group on
itself is generated by the right-invariant vector fields which we
denote by $\hat D_{IJ}$. These vector fields satisfy the following
constraints

\bea \hat D_{IJ}&=&\hat D_{JI}\nn \\ \bar{\hat
D}^{IJ}&=&\h^{IK}\h^{JL}\hat D_{KL} \eea

The algebra of the right-invariant vector fields is

\be
[\hat D_{IJ},\hat D^{KL}]=-4\d_{(I}{}^{(K}\hat D_{J)}{}^{L)} \ee

and they act on $USp(2n)$ by

\be
\hat D_{IJ} U_K{}^k=-\h_{IK} u_J{}^k-\h_{JK} u_I{}^k \ee

The $\hat D_{IJ}$'s split into $\hat D_{rs}, \hat D_{r's'}$ and
$\hat D_{rs'}$, the latter set corresponding to the isotropy
algebra $\gu(n)$ while the former two sets correspond to the coset.
We now define

\bea D_o&=& \h^{rs'} \hat D_{rs'}\nn \\ D_{rs'}&=& \hat
D_{rs'}-{1\over n}\h_{rs'}D_o \eea

and drop the hats on $D_{rs}$ and $D_{r's'}$. $D_o$
is the $\gu(1)$ derivative while the operators $D_{rs'}$ span the
$\gs\gu(n)$
part of the isotropy algebra. The operators $D_{rs}$ transform as
a symmetric second-rank tensor under $SU(n)$ and can be regarded as
the components of the $\bar\del$ operator on the coset, while the
operators $D_{r's'}$ are the conjugate set. With these conventions
$u_r{}^i$ and $u_{r'}{}^i$ have $U(1)$ charges $+1$ and $-1$
respectively.

The CR structure on the whole space $\hat M=M\xz U(n)\bsh USp(2n)$
is spanned by $D_{\a r}$ and $D_{rs}$, with involutivity being
guaranteed by virtue of the fact that each derivative of this set
(anti)-commutes with any other in the set.

To complete the picture we need the notion of a real structure. In
the $(1,0)$ case we define a conjugation by combining the antipodal
map on the two-sphere with complex conjugation. For an equivariant
function $A$ on $SU(2)$ we define

\be
A(u)\mapsto \tilde A(u)=\overline{A(\e u)} \ee

where $\e$ is the usual $\e$-matrix. Fields which have an even
$U(1)$ charge can be real under this operation, which simply means
that such a field contains real representations of $SU(2)$ in a
harmonic expansion.

In the $(n,0)$ case there is a very similar operation defined on
equivariant fields by

\be
A(u)\mapsto \tilde A(u) = \overline{A(\h u)} \ee

Again a field can be real under this conjugation if it has even
$U(1)$ charge. Explicitly, $\tilde u:=\overline{(\h u)}$ is given by

\be
\ba{rcrcrcr} \widetilde{(u_r{}^i)}&=&\bar
u^{r'}{}_i&=&(u^{-1})_i{}^{r'}&=&\h^{r's}u_{si}\\
\widetilde{(u_{r'}{}^i)}&=&-\bar
u^r{}_i&=&-(u^{-1})_i{}^{r'}&=&-\h^{rs'}u_{s'i} \ea \ee

We shall now apply this formalism to the $(2,0)$ tensor multiplet.
We define

\be
W:={1\over 2}\e^{rs} u_r{}^i u_s{}^j W_{ij} \la{w} \ee

We claim that $W$ is analytic. Since $D_{rs} u_t{}^i=0$, $W$ is
clearly analytic on $F$. Grassmann analyticity follows because

\bea D_{\a r} W&=&{1\over2}\e^{st}u_s{}^ju_t{}^k u_r{}^i D_{\a
i}W_{jk}\nn
\\
&=&\e^{st}u_s{}^ju_t{}^k u_r{}^i\left(\h_{ij}\l_{\a
k}-\h_{ik}\l_{\a j}-{1\over2}\h_{jk}\l_{\a i}\right) \la{wg} \eea

Since $u_r{}^i u_s{}^j\h_{ij}=0$ we deduce immediately that $D_{\a
r}W=0$. We note as well that the reality of $W_{ij}$ ensures that
$W$ is real under the conjugation defined above.

The converse of this result is also true, namely,  a  real analytic
field $W$ of charge $2$ defined on $\hat M$ can be expressed in the
form of \eq{w} where the Minkowski superspace field $W_{ij}$ obeys
the constraints of the $(2,0)$ tensor multiplet discussed
previously. The first part of this comes from the Bott-Borel-Weil
theorem which, applied to this case, states that such an analytic
field $W$ must have a short expansion of the form of \eq{w}. Note
that this implies that $W_{ij}$ is symplectic traceless, since the
trace is automatically absent in \eq{w}. Reality then implies that
$W_{ij}$ is real, and G-analyticity implies that it must satisfy
the constraint \eq{3}. This is because $D_{\a i}W_{jk}$ will in
general decompose into a $4$ and a $16$ under $USp(4)$, but the
latter will not give zero if substituted into the right-hand side
of \eq{wg}.

It is obvious that we can use $W$ to generate a family of superfields
$A_p:=W^p$ which are analytic on $\hat M$ and which have $U(1)$
charge $2p$ \cite{h}. The field $A_2$ is particularly important as it is the
supercurrent, which in Minkowski superspace has the form $W_{ij} W_{kl}|_{14}$ \cite{hsit}. The fact that only the
$14$-dimensional representation of $W\xz W$ appears in the product
is due to the properties of the $u's$. Indeed, in $SO(5)$ notation
$W$ transforms as a vector and the powers of $W$ that appear in the
series $A_p$ fall into the symmetric traceless representations of
$SO(5)$. This is very similar to the $N=4$ Yang-Mills case, where
the leading components of the KK multiplets fall into symmetric
traceless representations of $SO(6)$. To show that these composite
fields are superconformal we now turn to a discussion of
superconformal transformations in six dimensions.


\section{Superconformal transformations in Minkowski superspace}

Superconformal transformations in $D=6, (n,0)$ Minkowski
superspaces for $n=1,2$ have been discussed in refs \cite{p,gm}. The
situation is rather similar to four dimensions (see \cite{hh1} for a review). An infinitesimal
superconformal transformation of Minkowski superspace  can be
defined to be an infinitesimal diffeomorphism which preserves the
odd fermionic tangent bundle spanned by the vector fields $D_{\a
i}$. Dually, it can be defined to be an infinitesimal
diffeomorphism which preseves the even cotangent bundle spanned by
the one-forms $E^a:=dx^a-{i\over2}d\th^{\a
i}(\c^a)_{\a\b}\h_{ij}\th^{\b j}$.

Let $V$ be the vector field which generates such a diffeomorphism,
then we have

\be
 <D_{\a i}, \cL_V E^a>=0
 \la{sconf}
 \ee

where $<,>$ denotes the standard pairing between one-forms and
vector fields. If

\be
V=F^a\del_a + \vf^{\a i}D_{\a i} \ee

we find that \eq{sconf} implies that

\be
D_{\a i}F^a + i(\c^a)_{\a\b}\vf^{\b}_i=0 
\la{sconc}
\ee

from which

\be
\vf^{\a i}={i\over6}(\c_a)^{\a\b}D_{\b}^i F^a \ee

Thus a superconformal transformation is determined by a function
$F^a$ obeying the constraint

\be
D_{\a i}F^a={1\over6}(\c^{a}\c^b)_{\a}{}^{\b}D_{\b i} F_b \la{fcon}
\ee

It is not difficult to show that $F^a$ satisfying \eq{fcon}
contains precisely the parameters of the superconformal algebra.
Indeed, differentiating \eq{sconc} again one finds

\be
D_{\a i}\vf_{j}^{\b}=\d_{\a}{}^{\b}(f_{ij}+{1\over4}\h_{ij}\del_a
F^a) = \h_{ij} (\c^{ab})_{\a\b}L_{ab} \ee

where $f_{ij}=f_{ji}$ and $L_{ab}=\del_{[a}F_{ b]}$. One also finds
the conformal Killing equation

\be
\del_a F_b+\del_b F_a={1\over3}\h_{ab} \del^c F_c \ee

as well as the conformal Killing spinor equation

\be
\c_a\del_b \vf_i + \c_b\del_a\vf_i={1\over3}\h_{ab}\c^c\del_c\vf_i
\ee

If we define the components of $F$ in a $\th$-expansion to be
$\tilde F^a:=F^a|$, $\tilde\vf^{\a i}:=\vf^{\a i}|$ and $\tilde
f_{ij}:=f_{ij}|$, where the vertical bar denotes evaluation of a
superfield at $\th=0$, then $\tilde F^a$ is a conformal Killing
vector on ordinary Minkowski space, $\tilde\vf^{\a i}$ is a
conformal Killing spinor which contains the $Q$ and $S$
supersymmetry parameters and $\tilde f_{ij}$, which is constant in
$x$, is the $USp(2n)$ parameter.

It is easy to see that the commutator of two vector fields
satisfying \eq{sconf} is a vector field of the same type so that the
superconformal algebra can be expressed in terms of $F^a$ in the
form

\be
[F,F']=F''
\ee

where

\be
F''^a=F\cdot\del F'^a-F'\cdot\del F^a -i\vf^{\a
i}(\c^a)_{\a\b}\vf'^{\b}_i \ee

is easily checked to satisfy \eq{fcon} provided that $F$ and $F'$
do.

\section{Superconformal transformations on analytic fields}

The superconformal algebra is represented on $M\xz USp(2n)$ by
vector fields $V$ of the form

\be
V=V_o+V_u \ee

where $V_o$ is a superconformal Killing vector on $M$ and

\be
V_u= -{1\over2}f^{IJ}\hat D_{IJ}
\ee

with

\be
f^{IJ}=D_{\a}^{(I}\vf^{\a J)}
\ee

In addition we use the standard convention that $USp(2n)$ indices
are converted from lower case to upper case by means of $u$.

This vector field has the following properties: it defines a
representation of the superconformal algebra; it commutes with
the right-invariant vector fields on $USp(2n)$, and it preserves
CR-analyticity modulo isotropy group derivatives. Thus Grassmann analyticity is preserved for H-analytic fields because

\be
[D_{\a r},V]\sim D_{\a r}, D_{rs}
\ee

modulo isotropy group derivatives. These isotropy group terms have
the effect that a transformation of the above type, acting on a
G-analytic superfield $A$, say, with $U(1)$ charge $p$, does not
preserve G-analyticity so that a correction term is required. The
only possibility is to include a term proportional to $\del\cdot
F$, since this function is independent of $u$ and hence harmonic
analytic. The required transformation law is

\be
\d A=VA + {p\over3n}(\del\cdot F)A
\ee

It can be checked that this rule does preserve G-analyticity and
harmonic analyticity and that it does define a representation
because

\be
V_F (\del\cdot F')-V_{F'} (\del\cdot F)=\del\cdot [F,F']
\ee

Equivalently we can write

\be
\d A=\cV A + {p\over n(3-n)}\D A
\ee

where

\be
\cV=F\cdot\del + \vf^{\a r'}D_{\a r'}-{1\over2}f^{r's'}D_{r's'}
\ee

and

\bea
\D&=&\del\cdot F -D_{\a r'}\vf^{\a
r'}-{1\over2}D_{r's'}f^{r's'}\nn\\
&=&(3-n)\left({1\over3}\del\cdot F + \h^{rs'}f_{rs'}\right)
\eea

The vector field $\cV$ also gives a representation of the
superconformal algebra and has the property that it preserves
G-analyticity and CR analyticity up to isotropy group terms. The
function $\D$ is the divergence of
$\cV$. Note that there is no singularity for $n=3$ as the two
$(3-n)$ factors cancel from $\D$ and the transformation rule.

For the $(2,0)$ tensor multiplet field strength $W$ the $U(1)$
charge is 2 and hence we expect that

\be
\d W= V W + {1\over3}\del\cdot F W=\cV W + \D W
\la{vw}
\ee

To verify that this is indeed the case one starts from the transformation rule of the three-form $F$ which is

\be
\d F=\cL_V F
\la{vf}
\ee

Since

\be
F_{\a r\b s c}=-i(\c_c)_{\a\b}\e_{rs} W
\ee

we can find the transformation rule for $W$ by examining this component of \eq{vf}. It is straightforward to verify \eq{vw} by this means.


\section{Correlation functions}

In this section we consider the constraints which correlation
functions of analytic operators in $D=6, (2,0)$ should satisfy.
Even though we do not know the interacting theory it is still
possible to examine the Ward Identities of the theory. The
technique to be used here is the same one that has been employed to
study the correlation functions of analytic operators in $D=4$
superconformal field theories in a series of papers \cite{hw,hsw,ehssw,ehpsw,ehsw}, although we
are not able, in this case, to perform checks in harmonic
superspace perturbation theory. Indeed, even though one can present the (2,0) tensor multiplet in (1,0) superspace as a tensor multiplet together with a hypermultiplet, the (1,0) tensor multiplet does not admit an off-shell superfield formulation due to self-duality so that there there is no (1,0) superspace action even for the free theory. There have been several studies of 3-point functions in the literature, mainly for scalars \cite{3pt} or energy-momentum tensors \cite{em}, including some discussion of anomalies.

We denote an $N$-point correlation function of analytic operators
with charges $p_1,\ldots p_n$ by $<p_1\ldots p_N>$. The Ward
identity for such a correlation function reads

\be
\sum_{i=1}^N\left(\cV_i + \frac 13 p_i\D_i \right)<p_1\ldots
p_N>=0
\la{wi}
\ee

where we assume that all the points are separated.

The basic building blocks in the analysis of the Ward identities
are the propagator, which is the two-point function of two free
$W$'s, and the invariants. The propagator can be written

\be
<W(1)W(2)>\sim g_{12}={(12)\over x_{12}^4}
\ee

where

\be
(12):= u^{ij}(1)u_{ij}(2)
\ee

with

\be
u^{ij}:={\frac 12}\e^{rs} u_r{}^i u_s{}^j
\ee

The spacetime variable $x$ is modified in order to satisfy
Grassmann analyticity, and $x_{12}=x_1-x_2$ denotes the difference as usual. The analytic $x$ is related to the standard spacetime coordinate $x_S$ by

\be
x^{a}=x_S^a- {i\over2} \th^{\a r}(\c^a)_{\a\b}\th^{\b s'}\h_{rs'}
\ee

Given the propagator it is straightforward to write down the two-
and three-point functions. They are

\be
<p_1 p_2>\sim \d_{p_1 p_2} (g_{12})^{p_1}
\ee

and

\be
<p_1 p_2 p_3>\sim (g_{12})^{p_{12}} (g_{23})^{p_{23}} (g_{12})^{p_{23}}
\ee

where

\be
p_{ij}:={\frac 12}p_i+p_j-p_k,\qquad k\neq i,j
\ee

This formula for the three-point functions is rather similar to the one for $D=4, N=4$ Yang-Mills theory \cite{hsw}. Note that it gives the functional form of all of the three-point functions for all of the component operators which are related by AdS/CFT to the KK multiplets of $D=11$ supergravity on $AdS_7\xz S^4$. 

For more than three points, using standard conformal field theory
techniques, we can write  the correlation functions, or at least
those which do not vanish at leading order, as prefactors times
functions of invariants, where the prefactors, which are
constructed from the propagators, are solutions in the free theory.
For a given correlator the prefactor will satisfy the full Ward
identity \eq{wi}, but in general will not be unique. We thus have,
schematically,

\be
<p_1\ldots p_N>\sim P \xz F
\ee

where $F$ is a function of invariants.

It would be interesting to study, for example, four-point functions
using similar techniques to those of \cite{ehpsw} to see whether there can be non-trivial correlation functions even
though one does not know the interacting theory. We can make a few
speculations about certain special correlators, however, for which the
charges are chosen in a special way.

We recall that in four dimensional SCFT there are certain extremal and next-to-extremal correlators whose functional form is free and which obey non-renormalisation theorems \cite{dfmmrext,arufrov,biakov,ehssw,erdPer,arufrov2,defpv,ehsw}. It has recently been proposed that similar simplifications should occur in the present context \cite{d'h}.
As in four dimensions the prefactors are uniquely determined for
extremal correlators where the charge at point 1, say,  is equal to the
sum of the charges at the other points, $p_1=\sum_{i=2}^N p_i$. The
prefactor, $P$, is

\be
P=\prod_{i=2}^N (g_{1i})^{p_i}
\ee

In the analysis for such correlators in four dimensions in analytic superspace \cite{ehssw,ehsw} it was
found to be
helpful to use the reduction formula \cite{i} which relates $N$-point
functions, differentiated with respect to the coupling, to $(N+1)$-point functions with one integrated insertion
of the supercurrent. This trick is not available here, but we can
still attempt to make a straightforward analysis for four points. There are no nilpotent invariants for four points, essentially because one has enough supersymmetries, 32, to transform away all of the odd coordinates, so that the invariants are the supersymmetric completions of the zeroth-order invariants. There are two independent spacetime cross-ratios and two independent internal cross-ratios and the supersymmetric completions of all of these have singularities in the internal space which occur in, say, $y_{13}$ and $y_{24}$ together, where we use $y$ to denote the complex coordinates of the internal space.\footnote{The internal space can be thought of as complexified compactified 3-dimensional Minkowski space.} It is difficult to see how such singularities can be cancelled by the prefactors for extremal correlators in which each factor involves point 1. Although this is not a rigorous argument, it is very suggestive, and lends support to the conjecture recently made in \cite{d'h}.

Another way of looking at this problem would be to work in $(1,0)$ harmonic superspace and study the simpler problem of extremal
four-point
functions of hypermultiplet composites. The internal space in this case is the same as in $D=4, N=2$ harmonic superspace, i.e. $\bbC P^1$, so that the analysis resembles that given for extremal correlators in $D=4, N=2$ given in \cite{ehssw}.

For more than four points there will be nilpotent invariants exactly as in the four-dimensional case. These invariants are rather complicated, but they are all singular. It is therefore very plausible that $n$-point extremal correlators are also free \cite{d'h}. It should be possible to make these arguments using invariants rigorous by studying the singularities explicitly; one might also hope to find strong constraints on next-to-extremal correlators as in four dimensions \cite{ehssw,ehsw}. This has also been proposed in ref \cite{d'h}.

 \end{document}